\documentclass[a4paper,12pt]{article}

\usepackage{amsfonts}
\usepackage{graphicx}
\usepackage{amssymb}
\usepackage{amsmath}
\usepackage{feynmf}
\usepackage{cancel}

{\small}

\newcommand{\slashchar}[1] {\cancel{#1}}
\newcommand{\emiss}{\slashchar{E}}
\newcommand{\ptmiss}{\slashchar{p}}

\usepackage{fancyhdr}
\begin{document}
\begin{titlepage}
\begin{flushleft}
{\Large\bf TeV Mini Black Hole Decay at Future Colliders}
\end{flushleft}

\vspace{0.3cm}

\begin{flushright}
\parbox[t]{13cm}{
{\bf Alex Casanova\dag, Euro Spallucci\ddag}\\[4pt]
{\footnotesize\dag Dipartimento di Fisica Teorica, Dipartimento di Fisica, Universit\`a di Trieste and INFN, Sezione di Trieste}

\vspace{0.5cm}

{\footnotesize
{\bf Abstract}. It is generally believed that mini black holes decay by emitting elementary
particles with a black body energy spectrum. The original calculation lead to
the conclusion that about the $90$\% of the black hole mass is radiated away 
in the form of photons, neutrinos and light leptons, mainly electrons and 
muons.\\
With the advent of String Theory, such a scenario must be updated by including
new effects coming from the stringy nature of particles and interactions.
The main modifications with respect to the original picture of black hole
evaporation come from recent 
developments of non-perturbative String Theory globally referred to as  
\textit{TeV-Scale Gravity}. By taking for granted that black holes can be
produced in hadronic collisions, then their decay must take into account that:
(i) we live in a $D3$-Brane embedded into an higher dimensional bulk 
spacetime; (ii) fundamental interactions, including gravity, are unified 
at TeV  energy scale.
Thus, the formal description of the Hawking radiation mechanism has to be 
extended to the case of more than four spacetime dimensions and
include the presence of D-branes. This kind of topological defect in the
bulk spacetime fabric acts as a sort of ``~cosmic fly-paper~'' trapping
Electro-weak Standard Model elementary particles in our $(3+1)$-dimensional
universe. Furthermore,  unification of fundamental interactions at an energy
scale many order of magnitude lower than the Planck energy  implies that
any kind of fundamental particle, not only leptons, is expected to be emitted.\\
A detailed understanding of the new scenario is instrumental for  optimal
tuning of detectors  at future colliders, where, hopefully, this exciting new
physics will be tested.\\
In this article we review higher dimensional black hole decay, considering 
not only the emission of particles according to Hawking mechanism,  but also
their near horizon $QED/QCD$ interactions. The ultimate motivation is to build
up a phenomenologically reliable scenario, allowing a clear experimental
signature of the event.
}}
\end{flushright}

\vspace{1cm}

\noindent
{\footnotesize Submitted to: \textit{Class. Quantum Grav.}}

\vspace{2cm}

\noindent
{\footnotesize
\dag\; e-mail address: \texttt{casanova@infis.univ.trieste.it}\\
\ddag\; e-mail address: \texttt{spallucci@trieste.infn.it}}

\end{titlepage}

%%%%%%%%%%%%%%%%%%%%%%%%%%%%%%%%%%%%%%%%%%%%%%%%%%%%%%%%%%%%%%%%%%%%%%%%%%%%%%%%
\section{Introduction}

Recent developments, both in experimental and theoretical high energy physics,
seem to ``conjure'' the astonishing possibility to produce a mini black hole
in laboratory. From one hand, the next generation of particle accelerators,
as LHC, will  reach a center of mass energy of  $O(10)$ TeV; on the other
hand, non-perturbative String Theory, including $D$-branes, can accommodate 
\textit{large} extra dimensions characterized by a compactification scale
many order of magnitude bigger than the Planck length. A promising realization
of these ideas is  represented by ``TeV-Scale Gravity'' where it is possible
to introduce a unification scale, $M_*$, as low as some TeV (\cite{dvali:uno},
\cite{dvali:due}). If these theoretical models are correct
new exciting phenomenology (\cite{kc:origin}, \cite{hossen:uno})
is going to be observed at future colliders, as the physics of unified
interactions, including quantum gravity, will become accessible. No doubts, the
most spectacular expected event is a mini black hole production and
decay \cite{giddings:guida}. It has been conjectured
(\cite{thorne:hoop}, \cite{iop:rot}) that two partons
colliding with center of mass energy $\sqrt{\hat{s}}$ and impact parameter
smaller than the effective black hole event horizon radius $r_H(\sqrt{\hat{s}})$
will gravitationally collapse. One expects that proton-proton scattering at
LHC will produce, among many possible final states,
a rotating, higher dimensional, black hole \cite{myers:perry} with mass
$M_{BH}=\sqrt{\hat{s}}\sim 10$ TeV ; the estimated rate of black hole
production in geometrical approximation
(\cite{eardley:prod}, \cite{nambu:prod}, \cite{rychkov:prod}), is as high as one
per second (\cite{bh:lhc}, \cite{landsberg:due}),
less optimistic evaluations give $10^2-10^3$
per year. The evaporation mechanism considers mini black hole as
semi classical objects
%%%%%%%
\footnote{This is strictly valid only for black holes with masses $M_{BH}$
much above the fundamental higher dimensional Planck scale $M_*$
(see \cite{cav:due}).},
%%%%%%%
emitting particles with a black body energy spectrum \cite{art:hawk}
%%%%%%
\footnote{Throughout this paper we use the natural units $\hbar=c=k_B=1$.}:
%%%%%%

\begin{equation}\label{eqn:spettroHawk}
\langle\, N\,\rangle_{\omega m s}=
\frac{\vert\, A\,\vert^2}{e^{\left(\, \omega-m\,\Omega_H\,\right) /T_{BH}}
-(-1)^{2s}}\; .
\end{equation}
Here, $\langle\, N\,\rangle_{\omega m s}$ is the average number of particles
with energy $\omega$, spin $s$,  third component of angular momentum $m$,
emitted by a black hole with angular velocity $\Omega_H$ and temperature 
$T_{BH}$. $|A|^2$ is the ``greybody factor''
accounting for the presence of a  potential barrier surrounding the black hole.
All the relevant quantities (cross sections, decay rates, etc.)
are compute in the framework of relativistic quantum field theory,
which provides a low energy effective
description of black holes and particle interactions. From this point of view,
String Theory appears to provide ``only'' the general scenario, including extra
dimensions and $D$-branes, rather than computational techniques. On the other
hand, String Theory introduces the concept of \emph{minimal length} as the
minimal distance which can be physically probed
(\cite{euro:minlength1},\cite{euro:minlength2}). Including such
feature in the dynamics of black hole evaporation leads to a significant
change in the late phase of the process, as we shall discuss in the conclusion
of the paper.\\
The paper is organized as follows:
in Section \ref{sez:bhd} we recall the different phases of black hole 
evaporation process, mainly focusing on Schwarzschild black hole decay.
In Section \ref{sez:qced} we consider the QED/QCD interaction among particles
directly emitted by black holes via Hawking mechanism;
thus, in Subsection \ref{sez:phch} we report some numerical results
about formation and development of photosphere and chromosphere
around a Schwarzschild black hole.
In Section \ref{sez:ult} we analize in more detail parton fragmentation
into hadrons and black hole emergent spectra: in Subsection \ref{sez:Hpost}
we consider the case in which a chromosphere forms and develops,
while in Subsection \ref{sez:Hdirect} the case in which simple near horizon
parton hadronization occurs.
In Section \ref{sez:conclu} we conclude with a brief summary of the results
and some open problems.

%%%%%%%%%%%%%%%%%%%%%%%%%%%%%%%%%%%%%%%%%%%%%%%%%%%%%%%%%%%%%%%%%%%%%%%%%%%%%%%%
\section{Black Hole Decay}\label{sez:bhd}
By taking for granted that a mini black hole can be produced in hadronic
collisions, it will literally ``explode'' much before leaving any kind of
direct signal in the detectors
%%%%%%%%
\footnote{For an alternative scenario see \cite{casadio:uno}.}.
%%%%%%%%
From a theoretical point of view we can distinguish three main phases 
in the decay  process:

\begin{enumerate}
\item {\bf Spin Down phase}: during this early stage the black hole 
loses most of its angular momentum but only a fraction of its mass; 
numerical simulations \cite{page:due} indicate that
 $75$\% of  angular momentum and about $25$\% of  mass are radiated away.
  Thus, more than one half of the mass is emitted after the black hole
  has reached a non-rotating configuration.
\item {\bf Schwarzschild phase}: during this intermediate stage  a spherically
 symmetric, non-rotating, Schwarzschild black hole evaporates via Hawking
 radiation, losing most of its mass.
\item {\bf Planck phase}: this is the final stage of evaporation, when the
residual mass approaches the fundamental Planck scale $M_*$, i.e.  
$M_{BH}\simeq M_*$, and quantum gravity effects cannot be ignored.
\end{enumerate}

\noindent
A characteristic feature of higher dimensional models, like ``TeV-Scale
Gravity'',
 is that a $D$-dimensional black hole can emit energy and angular momentum
 both in our $(3 + 1)$-dimensional ``brane-universe'' (\emph{brane emission}), 
 and in the $D$-dimensional ``bulk''  where the brane is embedded 
 (\emph{bulk emission}) \cite{emparan:BvsB}. A simple estimate suggests that
 half energy is lost in the bulk \cite{cav:uno}; a more detailed
calculation
 shows that black holes radiate mainly on the brane, even if the ratio between
 energy emitted on the brane and in the bulk is not much greater than 
 one (\cite{kanti:finale},\cite{kanti:sca}).
As the Schwarzschild phase is considered the dominant stage,
and only the energy ($=$ particles) emitted on the brane can 
be directly observed,  we shall focus only on Schwarzschild black hole
brane emission, even if the black hole ``lives'' in  $D=(4+n)$ dimensions.
In this case, equation (\ref{eqn:spettroHawk}) reads:

\begin{equation}\label{eqn:sptHawknonrot}
\langle\, N\,\rangle_{\omega s}=
\frac{\vert\, A\,\vert^2}{e^{\omega /T_{BH}}-(-1)^{2s}}\; ,
\end{equation}
where

\begin{equation}\label{eqn:temp}
T_{BH}=\frac{n+1}{4\pi\, r_H}\; ,
\end{equation}
and $r_H$ is the event horizon radius \cite{kanti:finale} given by
\begin{equation}\label{eqn:radhor}
r_H=\frac{1}{\sqrt{\pi}\, M_*}\,
\left(\frac{M_{BH}}{M_*}\right)^\frac{1}{n+1}
\left(\frac{8\Gamma\left(\, \frac{n+3}{2}\,\right)}{n+2}\right)^\frac{1}{n+1}\;.
\end{equation}
In order to calculate power, $P$, and flux, $F$, emitted on brane
we proceed as follows.\\
First of all we calculate the ``greybody factor'' $\vert\, A\,\vert^2$
in equation (\ref{eqn:sptHawknonrot}). The ``greybody factor'' accounts for the
influence of spacetime curvature on particle motion; a particle created 
near the event horizon must cross a ``gravitational potential barrier''
in order to escape to infinity \cite{misner:gravitation}.
 Regarding brane emission, the ``greybody factor'' can be calculated as a
 transmission factor across the potential barrier of the $(4+n)$-dimensional
 Schwarzschild metric projected on the brane:

\begin{equation} \label{met:schwBR}
ds^2=-\left[\ 1-\left(\, \frac{r_H}{r}\, \right)^{n+1}\, \right]\, dt^2+
\left[\, 1-\left(\, \frac{r_H}{r}\,\right)^ {n+1}\,\right]^{-1}dr^2+r^2\,
\left(\, d\theta^2 +\sin^2 \theta d\varphi^2\, \right)\;.
\end{equation}

\noindent
Once $|A|^2$ is known (\cite{kanti:sca}, \cite{kanti:fermvet}, 
\cite{cvetic:finn}), it is possible to define a greybody cross section
\cite{teott:gen} as:
\begin{equation} \label{eqn:ottgen}
\sigma^{(s)}_{gb}\left(\,\omega\,\right)=\frac{\pi}{\omega^2}\,(2j+1)!\,\vert\,
A\,\vert^2\; ,
\end{equation}
where $j$ is the total angular momentum of the particle emitted by black hole.

\begin{figure}[htb]
\begin{center}
\includegraphics[angle=270,width=7.5cm]{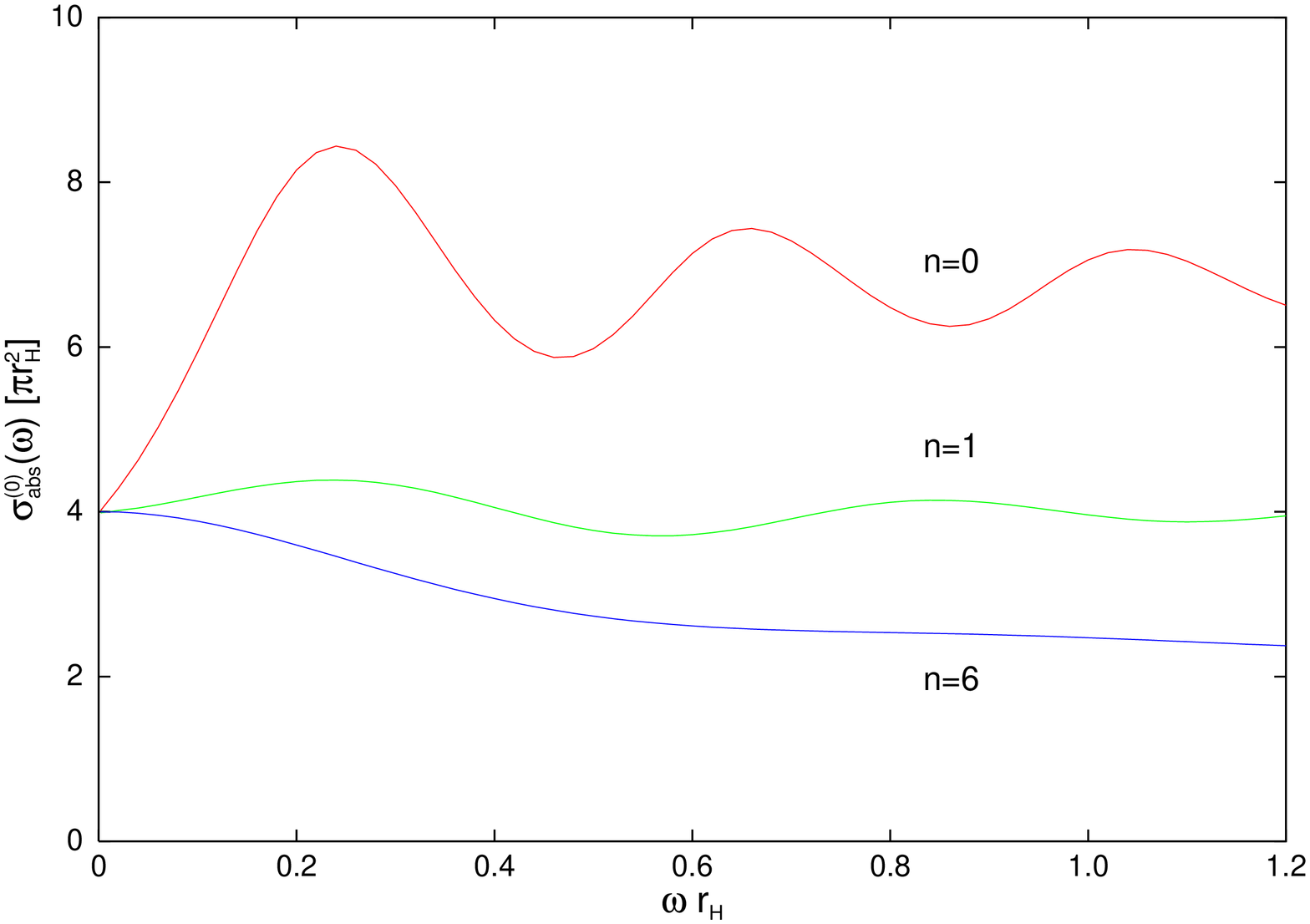}
\includegraphics[angle=270,width=7.5cm]{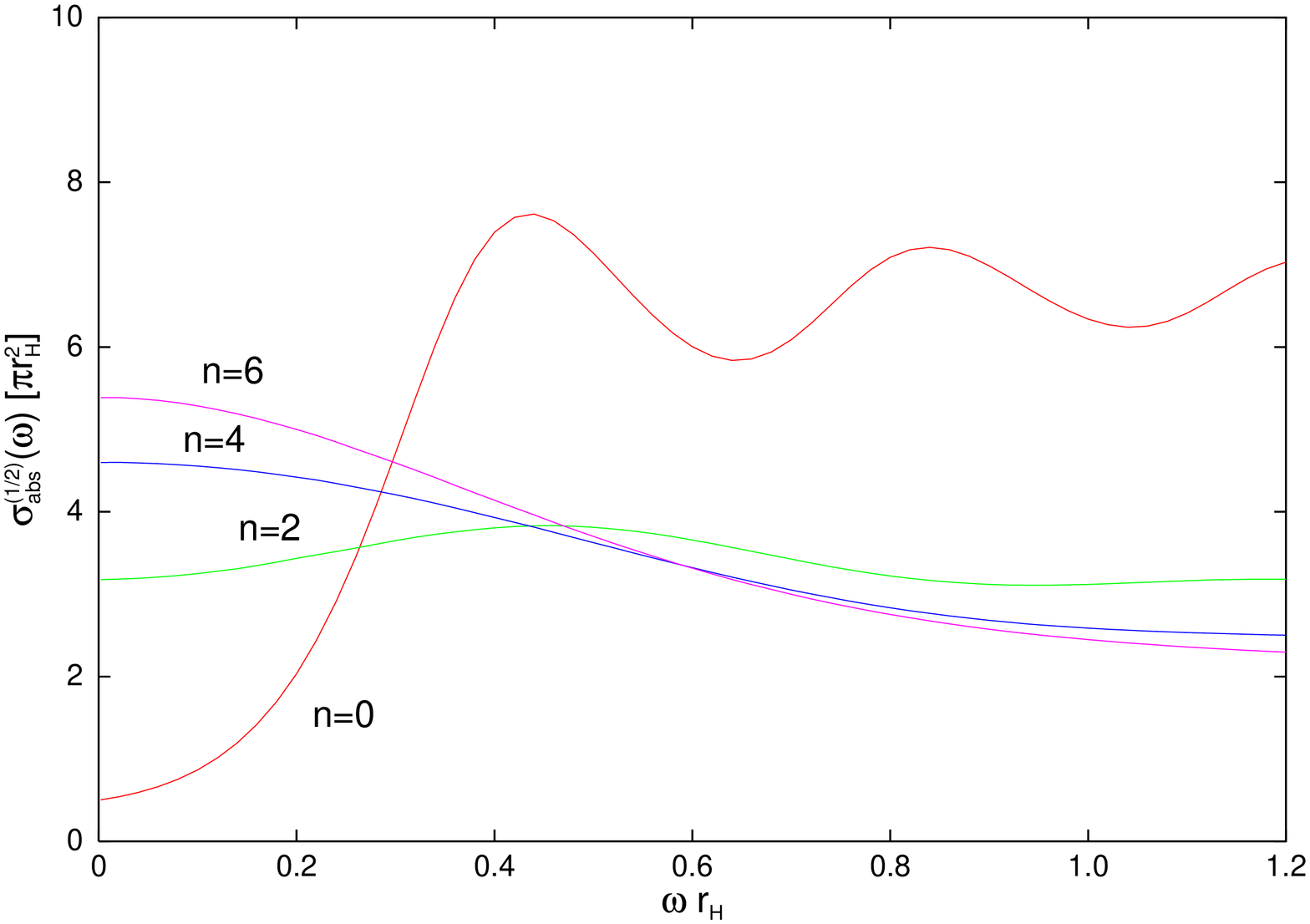}
\includegraphics[angle=270,width=7.5cm]{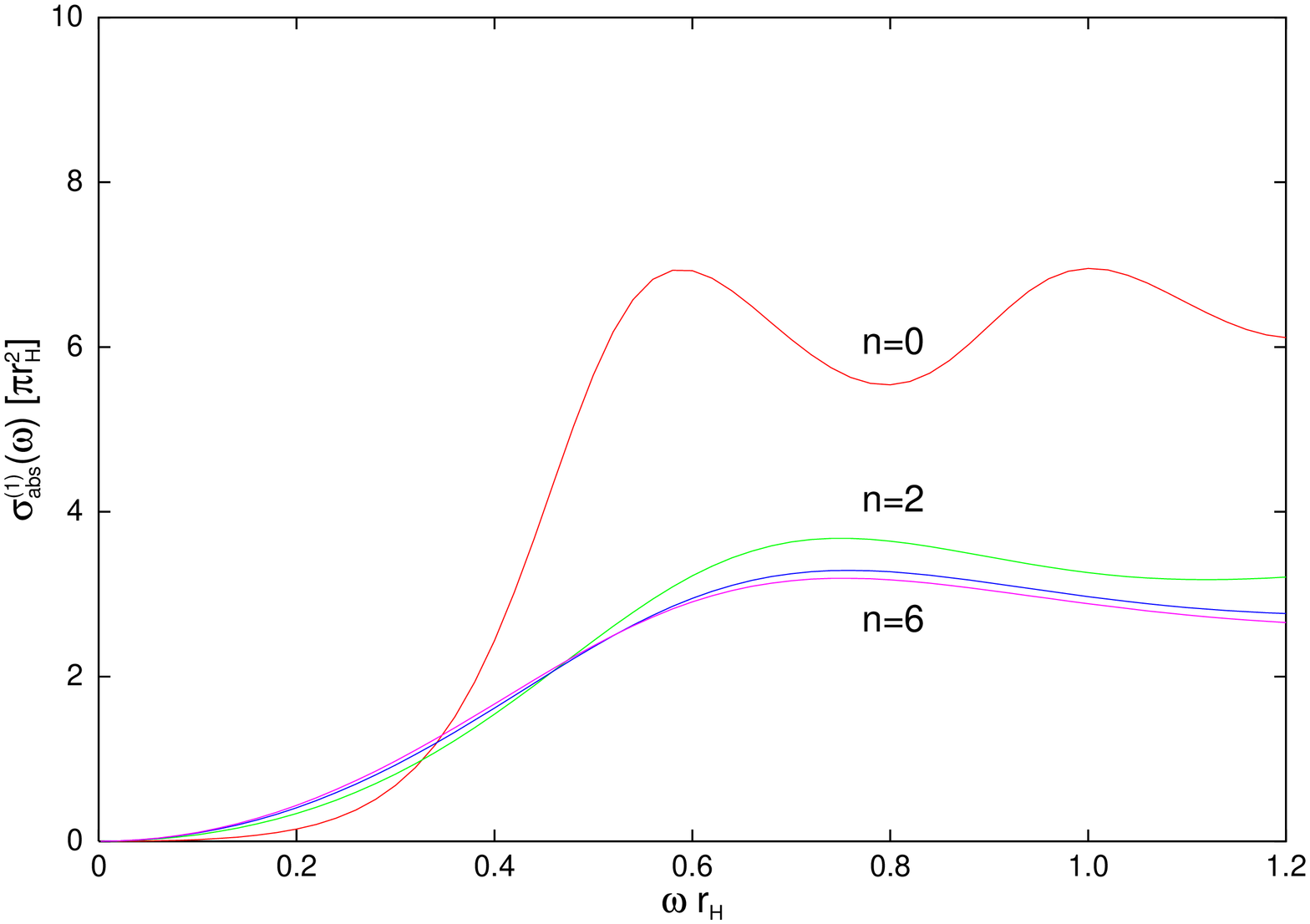}
\caption{\small Greybody cross section for brane emission from a
(4+n)-dimensional Schwarzschild black hole, for  with spin 0, 
$\frac{1}{2}$ and 1 particles (Reprinted figures with permission from
\cite{kanti:finale}. Copyright SISSA/ISAS 2003).}
\label{fig:GBF}
\end{center}
\end{figure}
\noindent
The results obtained in \cite{kanti:finale} is shown in Figure \ref{fig:GBF};
we can see that
\begin{itemize}
\item greybody factor is a function of particle energy ($\omega$) and spin
($s$);
\item greybody factor depends on the number $n$ of extra dimensions,
 which lead to conclude that it will be possible to ``measure'' the number of 
extra dimensions at future colliders (\cite{kanti:review}, \cite{harris:art2},
 \cite{rizzo:bhlhc});
\item asymptotically the greybody factor tends to geometric optics cross
section

\begin{equation}
\sigma_{g.o.}=\pi\left(\,\frac{n+3}{2}\, \right)^{2/(n+1)}\,
\left(\,\frac{n+3}{n+1}\,\right)\, r_H^2\; ,
\end{equation}

because of the finiteness of the gravitational potential barrier,
i.e. the potential barrier has a maximum of energy $\omega_{max}$ given by:
\begin{equation}
\omega_{max}= \frac{1}{r_H}\sqrt{\frac{n+1}{n+3}}\,\left(\,
\frac{n+3}{2}\,\right)^{-\frac{1}{n+1}}\;.
\end{equation}
\noindent
When a particle is emitted with an
energy $\omega\gtrsim\omega_{max}$ the black hole behaves as an ideal black
body
with area $\sigma_{g.o.}$.
\end{itemize}
By integrating equation (\ref{eqn:sptHawknonrot}) we can obtain the contribution
of spin $s$, Standard Model particle, to $P$ and  $F$:
\begin{itemize}
\item for $\omega \in [0,\omega_{max}]$, we  integrate thermal spectrum 
(\ref{eqn:sptHawknonrot}) with greybody factor $\sigma^{(s)}_{gb}(\omega)$;
\item for $\omega \in [\omega_{max},\infty]$, we  integrate thermal spectrum 
(\ref{eqn:sptHawknonrot}) with optical geometric cross section $\sigma_{g.o.}$.
\end{itemize}
Thus, we find

\begin{equation}
P=\left(\frac{dE}{dt}\right)=\frac{1}{2\pi^2}\, \int_0^{\omega_{max}}d\omega\,
\omega^3\,
\frac{\sigma^{(s)}_{gb}(\omega)}{e^{\omega/T_{BH}}-(-1)^{2s}}
+\frac{1}{2\pi^2}\,\int_{\omega_{max}}^\infty\,d\omega\,\omega^3\,
\frac{\sigma_{g.o.}\,  }{e^{\omega/T_{BH}}-(-1)^{2s}}\; ,
\end{equation}
and

\begin{equation}
F=\left(\frac{dN}{dt}\right)=\frac{1}{2\pi^2}\,\int_0^{\omega_{max}}d\omega\,
\omega^2\,
\frac{\sigma^{(s)}_{gb}(\omega)}{e^{\omega/T_{BH}}-(-1)^{2s}}+\frac{1}{2\pi^2}
\int_{\omega_{max}}^\infty d\omega\,\omega^2\, 
\frac{\sigma_{g.o.}}{e^{\omega/T_{BH}}-(-1)^{2s}}\;.
\end{equation}
Accounting for color and spin and considering an equal number of particles and 
antiparticles emitted by black hole, we find the power and flux emitted by a 
$(4+n)$-dimensional Schwarzschild black hole on brane through different type
of Standard Model particles, as reported in Table \ref{tab:PF}.

\begin{table}[htb]
\begin{center}
\begin{tabular}{|l||r|r|r|r|}
\hline {\rule[-3mm]{0mm}{8mm}}
\textsc{Power}  & $n=0$  & $n=2$  & $n=4$  & $n=6$  \\[6pt]
\hline
Quarks          & 64,9\% & 62,9\% & 61,8\% & 61,3\% \\
Charged Leptons & 10,8\% & 10,5\% & 10,3\% & 10,2\% \\
Neutrinos       &  5,4\% &  5,2\% &  5,2\% &  5,1\% \\
Photons         &  1,4\% &  1,7\% &  1,8\% &  1,9\% \\
Gluons          & 11,3\% & 13,5\% & 14,5\% & 14,9\% \\
Weak bosons     &  4,3\% &  5,1\% &  5,4\% &  5,6\% \\
Higgs           &  1,9\% &  1,0\% &  1,0\% &  1,0\% \\
\hline
\end{tabular}\\[10pt]
\begin{tabular}{|l||r|r|r|r|}
\hline {\rule[-3mm]{0mm}{8mm} }
\textsc{Flux}   & $n=0$  & $n=2$  & $n=4$  & $n=6$  \\[6pt]
\hline
Quarks          & 66,5\% & 64,1\% & 62,1\% & 60,8\% \\
Charged Leptons & 11,1\% & 10,7\% & 10,4\% & 10,1\% \\
Neutrinos       &  5,5\% &  5,3\% &  5,2\% &  5,1\% \\
Photons         &  1,2\% &  1,5\% &  1,8\% &  1,9\% \\
Gluons          &  9,3\% & 12,4\% & 14,1\% & 15,2\% \\
Weak bosons     &  3,5\% &  4,6\% &  5,3\% &  5,7\% \\
Higgs           &  3,1\% &  1,3\% &  1,2\% &  1,2\% \\
\hline
\end{tabular}
\caption{Power and Flux emitted by a (4+$n$)-dimensional Schwarzschild black 
hole on brane through Standard Model particles.}
\label{tab:PF}
\end{center}
\end{table}
\noindent
We see that $\approx 75$\% of  decay products are quarks, 
anti quarks and gluons, while only $\approx 12$\% are charged leptons and 
photons,
each particle carrying hundreds GeV of energy.
 We conclude that the black hole decay is dominated by partons; since they
 cannot be directly observed, we must to take into account fragmentation
 into hadrons. Therefore, in next section we are going to study the effects of
 $QED/QCD$ interactions (not only hadronization) among particles emitted near
 event horizon, in order to understand what kind of spectra we can expect to
 detect.

%%%%%%%%%%%%%%%%%%%%%%%%%%%%%%%%%%%%%%%%%%%%%%%%%%%%%%%%%%%%%%%%%%%%%%%%%%%%%%%%
\section{QED and QCD Effects}\label{sez:qced}

In previous section we have outlined the evaporation process of a mini black 
hole produced at colliders, with a main focus on the ``Schwarzschild Phase'',
concluding that brane black hole emission is dominated by quarks and gluons.
This emission, which we shall call ``\emph{direct emission}'', consists in the
near horizon creation of Standard Model particles through Hawking Mechanism
and in their flowing to infinity. However, the  picture consisting  of
Standard Model elementary
particles  freely escaping to infinity with a black body energy spectrum
must be somehow improved by including QCD interaction effects as black hole
emission is dominated by quarks and gluons. The inclusion of their interactions
is instrumental to build up phenomenologically reliable model predicting
the form  of the spectrum to be looked for.

In the next section we shall consider in more detail
parton fragmentation into hadrons, here we are going to discuss something
which could happen before hadronization, i.e. the appearance of a quark-gluon
plasma around the black hole (\cite{heckler:uno}, \cite{heckler:due}).
One starts by considering quarks and gluons emitted
according to a black body law. It follows that the parton density
 near the event horizon grows as $T_{BH}^3$. If temperature is high enough,
 i.e. above some critical value,
one expects from $QCD$ quarks and gluons to interact  through
 ``\emph{bremsstrahlung}'' and ``\emph{pair production}'' processes.
These reactions are $2\rightarrow 3$ body processes increasing the number
of quarks and gluons nearby black hole and leading to a kind of quark/gluon
plasma surrounding the event horizon. While  propagating through this plasma
quarks and gluons lose energy. When the average energy is low enough partons
fragment into hadrons.\\
Similar arguments can be applied to photons, electrons and positrons as well,
provided we replace $QCD$ interactions with the corresponding $QED$ processes. 
Thus, we can define two distinct regions surrounding the black hole as follows:
\begin{itemize}
\item {\bf Photosphere}, is the spatial region around the black hole where 
 ``QED-brems\-strahlung'' and ``QED-pair production'' lead to formation
 of an $e^\pm, \gamma$ plasma;
\item {\bf Chromosphere}, is the spatial region around black hole where
 ``QCD-bremsstrahl\-ung'' and ``QCD-pair production'' lead to formation of a
 quark/gluon plasma.
\end{itemize}
Spectrum distortion induced by the two ``atmospheres'' defined above will
be discussed in the remaining part of this section.

An analytic description  of photosphere and chromosphere dynamics is not
available at present; the best one can do is to resort to numerical resolution
of Boltzmann equation for the interacting particle
distribution function \cite{cline:num}. In order to get a first insight
of the phenomena we are considering, we shall introduce a further
simplification: we shall use ``flat spacetime'' diffusion relations even if
we are going to study the problem in the geometry of a Schwarzschild black
hole.\\
The number $ dn\left(\, \vec{p}\, ,r\, \right)$
of particles per unit volume element with momentum between $\vec{p}$ and
$\vec{p}+d\vec{p}$ is
$$
dn\left(\,\vec{p}\, ,r\,\right)=  f\left(\,\vert\, \vec{p}\,\vert\ ,r\,\right)\,
d^3p\,r^2 \, dr\, d\Omega\; ,
$$
where the distribution function $f\left(\,\vert\, \vec{p}\,\vert\ ,r\,\right)$
solves the Boltzmann equation
\begin{equation}\label{eqn:boltzrad}
\frac{\partial}{\partial r}f\left(\, p\ ,p_t\ ,r\,\right)=
\frac{1}{v_r}C[f\left(\, p\ ,p_t\ ,r\,\right)]\;.
\end{equation}
In (\ref{eqn:boltzrad}) we introduced the radial velocity $v_r$, the transverse
component of the momentum $p_t$ and defined $p\equiv|\,\vec{p}\,|$. $C[\, f\, ]$
is the \emph{collision term} built out of scattering cross sections encoding
information about interactions taking place inside photo and chromosphere.

According to QED (QCD), electrons and positrons (quarks and antiquarks) can lose
energy through the emission of photons (gluons):
$
e^\pm e^\pm\rightarrow e^\pm e^\pm\gamma \, (qq\rightarrow qqg).
$
Corresponding Feynman diagram is:
\begin{center}
\begin{fmffile}{brems}
\begin{fmfgraph}(80,80)
\fmfpen{thick}
\fmfbottom{i1,i2}
\fmftop{o1,o2,o3}
\fmf{fermion}{i2,v3,o3}
\fmf{photon}{v2,v3}
\fmf{fermion}{i1,v2,v1,o2}
\fmf{photon}{v1,o1}
\fmfdotn{v}{3}
\end{fmfgraph}
\end{fmffile}
\end{center}
In the ultra-relativistic limit, differential cross section in the center of
mass frame reads \cite{jauch:rohrlich}:
\begin{equation}\label{eqn:bremsdiff}
\frac{d\sigma}{d\omega}\approx\frac{\alpha^3}{E\omega m_f^2}
\left[\frac{4}{3}\left(E-\omega\right)+\frac{\omega^2}{E}\right]\times
\left[\log\left(\frac{4E^2(E-\omega)}{m^2_f\omega}\right)-\frac{1}{2}\right]\; ,
\end{equation}
where $m_f$ is the fermion mass, $\alpha$ is the electromagnetic (strong)
coupling costant, $E$ is the initial energy of each fermion, and
$\omega$ is the energy of emitted photon (gluon).\\
%The equation (\ref{eqn:bremsdiff}) implies that the probability to emit a
%photon diverges as its energy goes to zero; however, the emission of a zero
%energy photon has not effect on electron that emits it, since the electron
%will not lose energy. Therefore, either to eliminate the divergence
%in (\ref{eqn:bremsdiff}) or to account those photons which carry away
%significant energy,
To avoid infra-red divergences, it is convenient to use energy-averaged
total cross section (\cite{heckler:uno}, \cite{cline:num}):
\begin{equation}\label{eqn:brems}
\sigma\left(\,E\,\right) =
\int_0^\infty \frac{\omega}{E}\left(\, \frac{d\sigma}{d\omega}\,\right)
d\omega\approx \frac{\alpha^3}{m_f^2}\log\left(\frac{2E}{m_f}\right)\;.
\end{equation}
%\subsection{QED: Pair production}
The same type of cross section is obtained for pair production,
$
e^\pm\gamma\rightarrow e^\pm e^+ e^- \, (qg\rightarrow q\overline{q}q)
$; corresponding Feynman diagram is:
\begin{center}
\begin{fmffile}{pp}
\begin{fmfgraph*}(80,80)
\fmfpen{thick}
\fmfbottom{i1,i2}
\fmftop{o1,o2,o3}
\fmf{fermion}{i1,v2,v3,o3}
\fmf{fermion}{o1,v1,o2}
\fmf{photon}{v1,v2}
\fmf{photon}{i2,v3}
\fmfdotn{v}{3}
\end{fmfgraph*}
\end{fmffile}
\end{center}
At order $\alpha^3$ one finds:

\begin{equation}\label{eqn:pp}
\sigma\left(\,\omega\,\right)
 \approx\frac{\alpha^3}{m_f^2}\log\left(\frac{2\omega}{m_f}\right)\;,
\end{equation}
where $\omega$ is the incoming photon (gluon) energy.\\
Since $\sigma \propto 1/ m^2_f $, \emph{heavy fermions minimally affect
photo/chromosphere development}, and justify photo/chromosphere description
in terms of electrons, positrons and light quarks alone, as in \cite{cline:num}.

Scattering processes discussed above do not occur in vacuum, rather they
take place in an hot plasma of almost-radially moving particles, which is what
 we call photo/chromosphere. A simple way to take into account finite 
temperature effects consists in replacing  vacuum fermion masses 
($m_f$ in (\ref{eqn:brems}) and (\ref{eqn:pp})) with their thermal counterparts
\begin{equation}\label{eqn:mth}
m_{th}^2=m_f^2+M^2\left(\, T\,\right)\; ,
\end{equation}
where $T$ is the plasma temperature and $M$ is often referred to as the plasma 
mass.\\The position (\ref{eqn:mth}) gives the propagator pole in momentum space
with an accuracy of $10$\% \cite{weldon:temp}.
For finite temperature gauge theories, one finds
$$
M^2\left(\, T\, \right)=g^2\, C\left(\, R\, \right)\, \frac{T^2}{8}\; ,
$$
where $g$ is gauge coupling  constant, $C(R)$ is the quadratic Casimir invariant 
for gauge group representation $R$. Relevant values of $C(R)$ in numerical 
computations \cite{cline:num} are: $C(R)=1$ and $C(R)=4/3$ corresponding to
$U(1)$ and $SU(3)$ fundamental representation, respectively.
In the hot plasma scenario,  fermions move inside
photo/chromosphere  as ``free'' particles with a temperature dependent
effective mass $m_{th}$.\\
In next subsection we shall report main results for a numerical resolution of
Boltzmann equation (\ref{eqn:boltzrad}).

\subsection{Numerical Results}\label{sez:phch}

Let us start this subsection by investigating formation and development of
photo/chro\-mosphere around a  four-dimensional Schwarzschild black
hole \cite{cline:num}, then we shall discuss as photo/\-chromosphere features
depend from the number of extra-dimensions.

\subsubsection{Photosphere.}

We already mentioned that scattering processes become important only beyond
some critical temperature $T_c^{QED}$.\\
Let us introduce $\mathcal{N}(r)$ as the number of collisions a typical
particle undergoes between the event horizon and some larger radius $r>r_H$. 
$\mathcal{N}(r)$ can be written in terms of the bremsstrahlung and
pair production mean free path, $\lambda(r)$, as
$$
\mathcal{N}(r)=\int_{r_H}^r \frac{dr}{\lambda(r)}\;.
$$
We say that a photosphere surrounds the black hole if every particle is 
scattered at least once between the event horizon and infinity:
\begin{equation}\label{ct}
\lim_{r\rightarrow \infty} \mathcal{N}(r)\geq 1.
\end{equation}
Thus, the critical temperature $T_c^{QED}$ is the black hole temperature
giving one for the  limit (\ref{ct}). From numerical analysis we get
%%%%%%%%%%
\footnote{In \cite{heckler:uno}, $T_c^{QED}\simeq 45.2$ GeV.}:
%%%%%%%%%%
$$
T_c^{QED}\simeq 50\mbox{ GeV}.
$$

\begin{figure}[htb]
\begin{center}
\includegraphics[width=7.6cm]{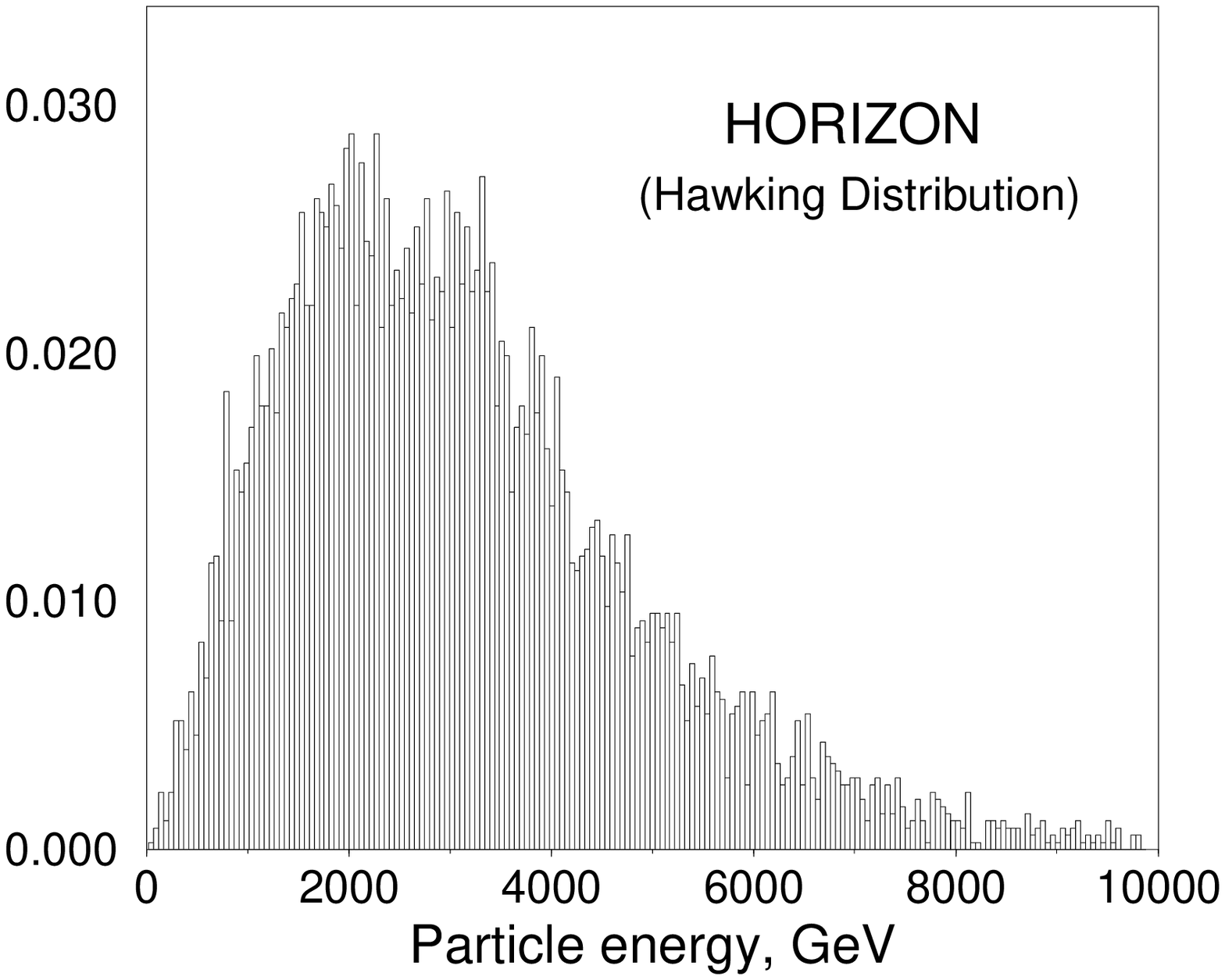}
\includegraphics[width=7.6cm]{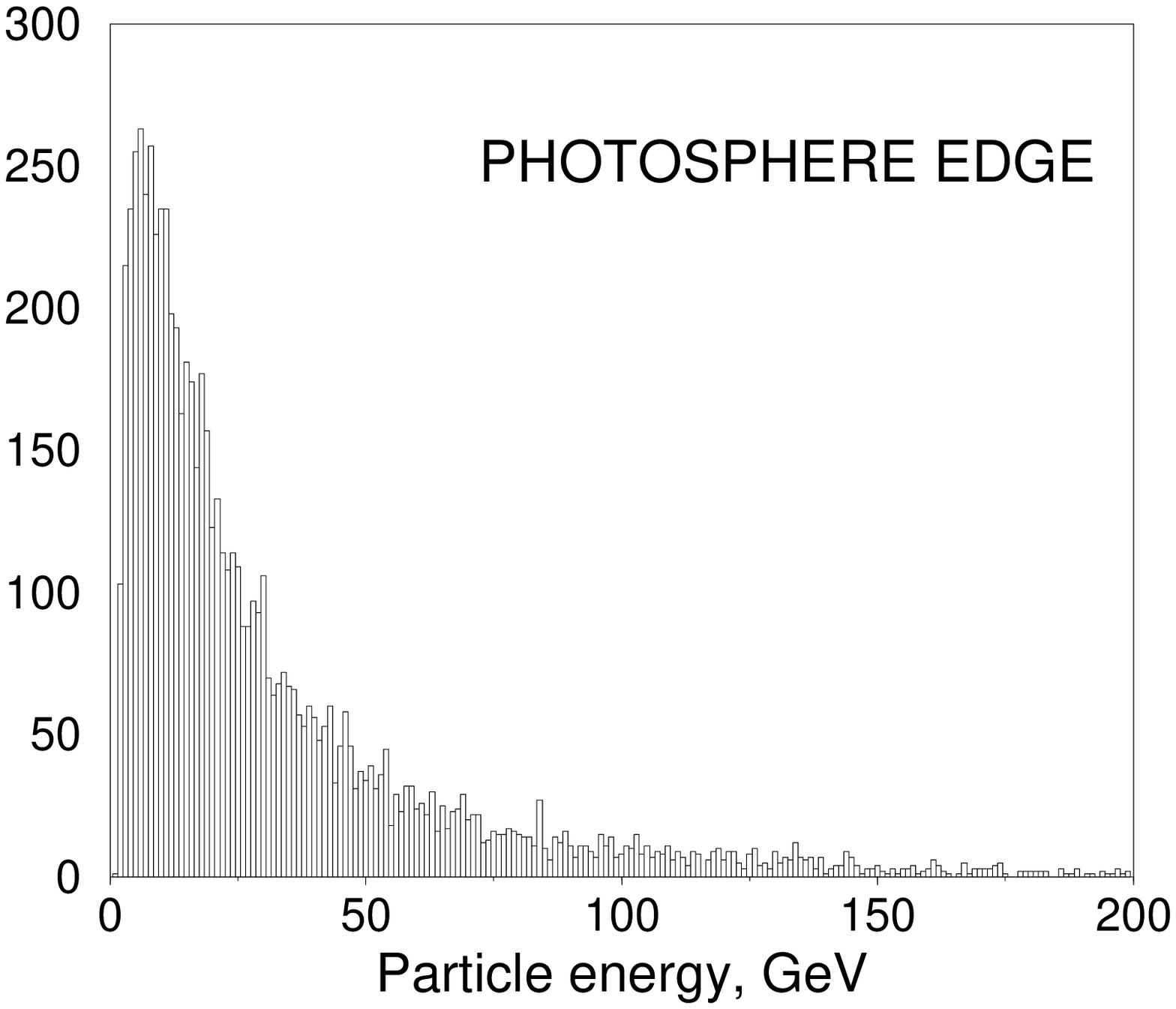}
\end{center}
\caption{\small Particle number versus energy for a 4D Schwarzschild
black hole with $T_{BH}=1000$ GeV. {\bf (a)\,:} near the horizon.
{\bf (b)\,:} at the outer photosphere surface
(Reprinted figures with permission from \cite{cline:num}.
Copyright 1999 by The American Physical Society).}
\label{fig:11}
\end{figure}

Another important photosphere parameter is the inner radius $r_i$ which can be
defined by $\mathcal{N}\left(\, r_i\, \right)=1$,
i.e. the mean radial distance for a particle to be scattered once.
\begin{figure}[htb]
\begin{center}
\includegraphics[width=7.6cm]{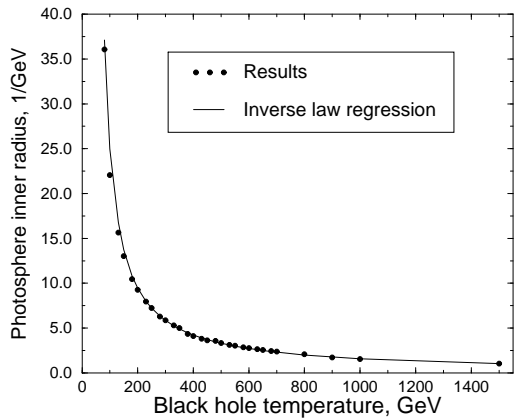}
\end{center}
\caption{\small Radii for inner photosphere surface for different temperatures;
remember that 1 GeV$^{-1}\simeq 0.197$ fm (Reprinted figure with permission
from \cite{cline:num}. Copyright 1999 by The American Physical Society).}
\label{fig:6}
\end{figure}
%{}
Data fit (Figure \ref{fig:6}) gives
\begin{equation}\label{eqn:inner}
\begin{array}{cc}
r_i=\frac{1}{\kappa T_{BH}},& \kappa=(6.446\pm 0.003)\times 10^{-4}.\\
\end{array}
\end{equation}
The inverse dependence from the temperature is due to the fact that 
bremstrahlung and pair production mean free path decreases with the temperature.
Thus, the inner edge of the photosphere is closer to
the event horizon.
By inserting $r_H=\frac{1}{4\pi T_{BH}}$ in (\ref{eqn:inner}) we get:
\begin{equation}\label{eqn:rirH}
r_i=\frac{4\pi}{\kappa}r_H\simeq 2\times 10^4 r_H.
\end{equation}
Even in the case $D\geq 5$, we would not expect a different behavior
for inner radius.
Indeed, bremsstrahlung and pair production are scattering processes involving
Standard Model particles bound to the $D3$-brane.\\
From an experimental point of view, it is important to know the
average final energy $\overline{E}_f$ of particles when they start propagating
without significant interactions, i.e. the particle average energy at outer
photosphere surface, because this is the expected energy to be eventually
released in a detecting device.
\begin{figure}[htb]
\begin{center}
\includegraphics[width=7.6cm]{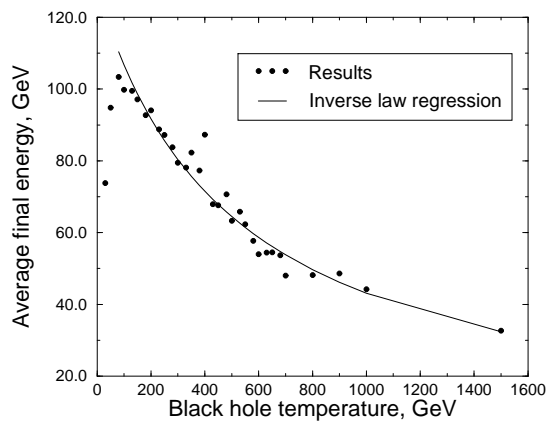}
\end{center}
\caption{\small Particle average energy at outer photosphere surface versus
temperature for $T_{BH}\leq 1.5$ TeV (Reprinted figure with permission from
\cite{cline:num}. Copyright 1999 by The American Physical Society).}
\label{fig:9}
\end{figure}
Figure \ref{fig:9} shows $\overline{E}_f$ for different black hole temperature
$T_{BH}$.
\begin{table}[htb]
\begin{center}
\begin{tabular}{|c||c|c|c|}
\hline {\rule[-3mm]{0mm}{8mm}}
$T_{BH}$ (GeV)        & 60     & 300    & 1000   \\
\hline {\rule[-3mm]{0mm}{8mm}}
$\overline{E}_f$ (GeV)& $97.0$ & $79.5$ & $44.2$ \\
\hline
\end{tabular}
\end{center}
\caption{\small Average final energy for some values of temperature
(Reprinted table with permission from \cite{cline:num}.
Copyright 1999 by The American Physical Society).}
\label{tab:e_f}
\end{table}
In Table \ref{tab:e_f} we  report some values extrapolated from
Figure \ref{fig:9}. Since, in the first approximation, particles are emitted
with a black body spectrum, the average energy of a particle close to the event
horizon is given by $\overline{E}_i\sim 3T_{BH}$; thus, we notice that for
$T_{BH}=60$ GeV, $\overline{E}_f$ is
not much different from $\overline{E}_i$,
because the temperature is only a little higher than
$T_c^{QED}$. In this regime the rate
of bremsstrahlung and pair production processes is not high enough to decrease
in a significant way particle energy. On the other hand, for $T_{BH}= 1000$ GeV,
$\overline{E}_f$
is much smaller than its initial value.
These results display in a clear way the role of the photosphere: above
a critical temperature the number of particles grows and the 
average energy decreases. This effect is  enhanced at higher black hole 
temperature. Figure \ref{fig:11} shows up this behavior in a quite evident way.
%{}

\subsubsection{Chromosphere.}

\begin{figure}[htb]
\begin{center}
\includegraphics[width=7.6cm]{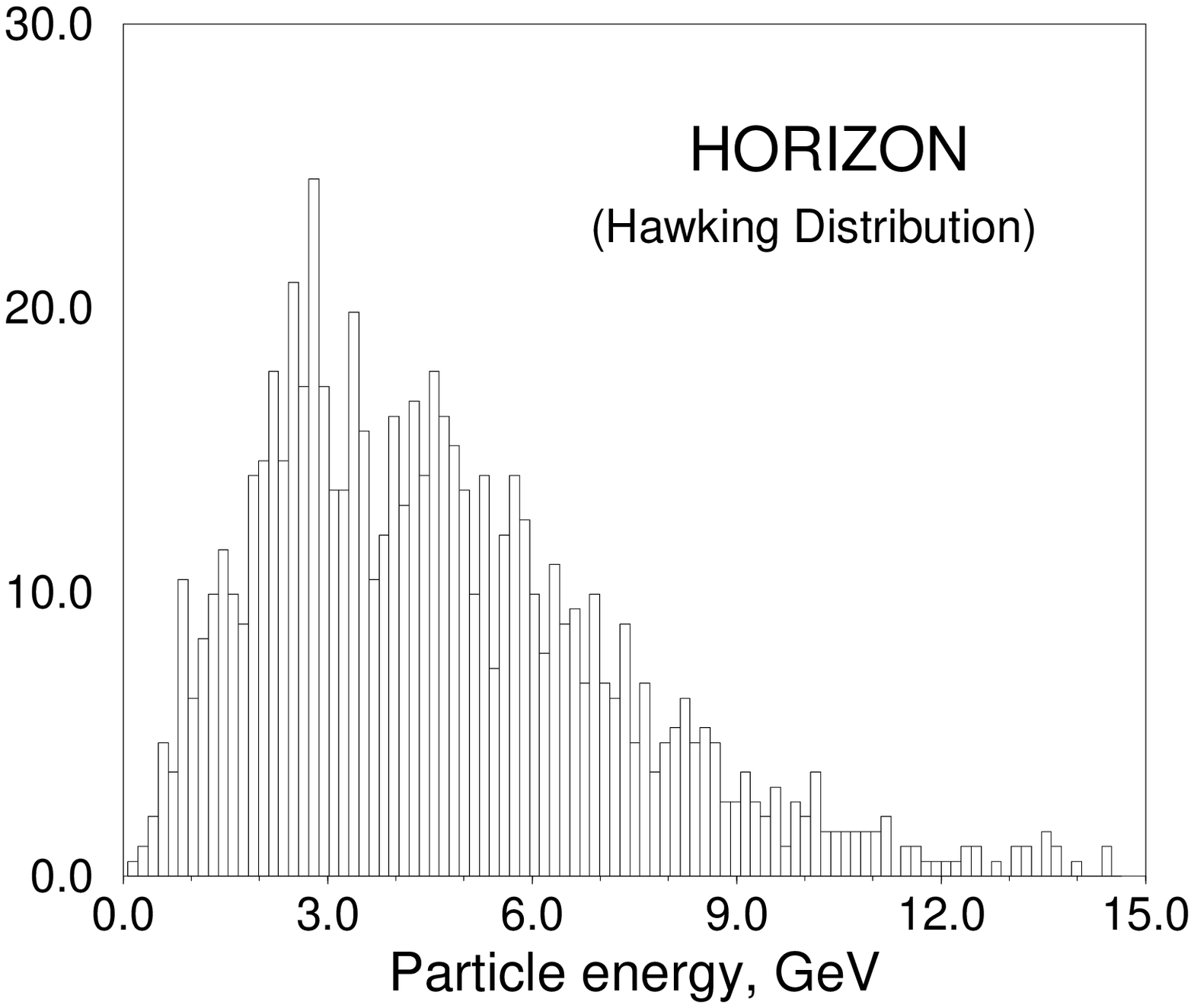}
\includegraphics[width=7.6cm]{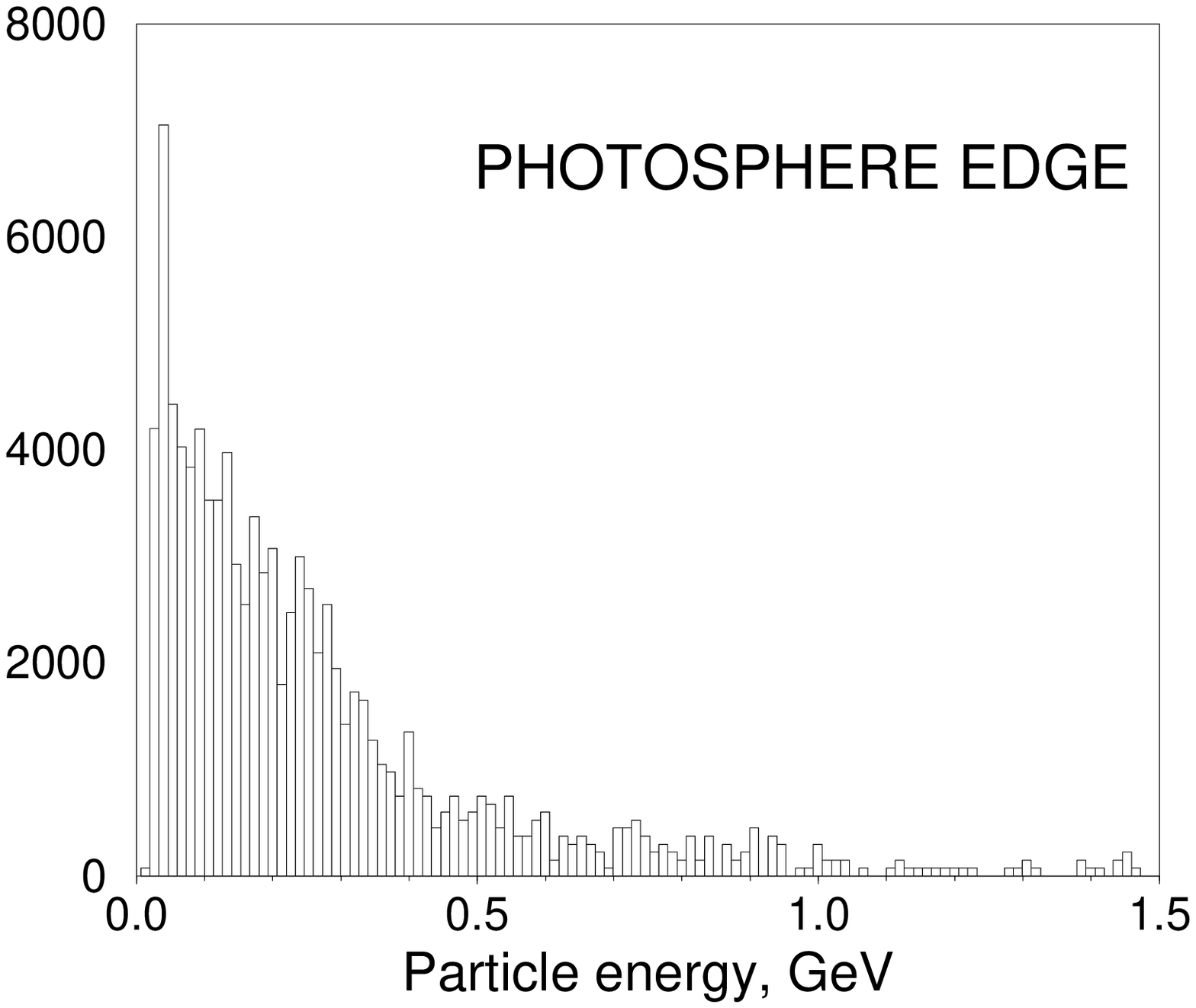}
\end{center}
\caption{\small Particle number versus energy for a 4D Schwarzschild black hole
with $T_{BH}=1.5$ GeV. {\bf (a)\,:} Hawking emission spectrum at the horizon
and {\bf (b)\,:} final spectrum emerging from chromosphere (Reprinted figures
with permission from \cite{cline:num}.
Copyright 1999 by The American Physical Society).}
\label{fig:18}
\end{figure}

By following the same procedure adopted for the photosphere, we find
$$
T_c^{QCD}\simeq 175\mbox{ MeV}.
$$
This result agrees with the analytic estimate in \cite{heckler:uno}. Thus,
 we conclude that whenever black hole temperature is high enough to produce
 a photosphere, then a chromosphere must be present as well. In such a case,
 the chromosphere inner radius is close to the horizon, i.e.
$r_i\sim r_H$: a black hole with temperature
$T_{BH}\gtrsim T_c^{QCD}$
 emits interacting quarks and gluons in the strong coupling regime
 with initial black body average energy $\overline{E}_i\sim 3T_{BH}$.
 As they propagate towards infinity their average energy decreases below
 $\Lambda_{QCD}$ and then fragment into hadrons. In this way,
one finds the average final energy for quarks and gluons to be
$$
\overline{E}_f\sim 200-300\mbox{ MeV}.
$$
Transition from partons into hadrons ``marks'' the position of chromosphere 
edge. Chromosphere emerging particle spectrum is fairly different from the direct
emission spectrum, as there are many more particles with a lower average energy
(Figure \ref{fig:18}).

\subsubsection{Concluding Remarks.}

Main results obtained in this section can be summarized as follows:
\begin{itemize}
\item the presence of a photosphere, or a chromosphere, 
surrounding the event horizon
implies a proliferation of emitted particles; energy conservation leads to a 
lower average energy per particle. Thus,  direct emission spectrum is modified:
a black hole with horizon temperature $T_{BH}=1.5$ GeV, effectively behaves
as black body  at  temperature  about 100 MeV (Figure \ref{fig:18}).\\
 When emitted particle interactions are properly accounted for,
the ``free'' Hawking spectrum is shifted to a an effective black body spectrum
to a temperature lower than the black hole temperature.
\item Whenever $T_{BH}\gtrsim \Lambda_{QCD}$, black hole emits quarks and
gluons and direct emission is partons dominated (Table \ref{tab:PF});
furthermore, since $T_c^{QCD}\lesssim \Lambda_{QCD}$,
strongly coupled  quarks and gluons form a chromosphere
surrounding the event horizon. Thus, one concludes
that final particle spectra will be dominated by hadrons,
coming from parton confinement, and their decay products.
\end{itemize}
Results reported in this section  have been obtained in the
geometry of a four-dimensional Schwarzschild black hole. However, examining
a more general $D$-dimensional case, the following features have to be taken
into account:
\begin{itemize}
\item scattering processes introduced previously are
 $D$-independent, because they involve only $3$-Brane confined Standard Model 
 particles; so cross sections (\ref{eqn:brems}) and (\ref{eqn:pp})
are unaffected;
\item brane emission is an intrinsically four-dimensional process, thus, 
 photo/chromo\-sphere dynamics, emitted power, emitted particle number,
 average initial
 energy, near horizon particle  density  don't change in higher dimensions;
\item brane emission is dominated by quarks and gluons independently of $D$,
furthermore flux percentages of partons and photons, electrons and
positrons, are quite constant.
\end{itemize}
All these considerations lead us to conclude that the four
dimensional picture of black hole decay does not significantly change
when moving to higher dimensions.

%%%%%%%%%%%%%%%%%%%%%%%%%%%%%%%%%%%%%%%%%%%%%%%%%%%%%%%%%%%%%%%%%%%%%%%%%%%%%%%%
\section{Hadronization and Emergent Spectra}\label{sez:ult}

We have seen in previous sections that final spectra, to be measured by
an asymptotic observer, are
going to
be dominated by hadrons, coming from parton confinement, and their decay
products,
mainly photons, neutrinos and $e^\pm$. For this reason, we are going to
consider first quarks  and gluons
hadronization
 and then black hole emergent spectra in two cases: i) a chromosphere forms
 and develops; ii) simple near
horizon
 parton hadronization occurs (as in \cite{macgibbon:webber}).

In order to recall some usefull formula to study hadronization and hadron
decay processes, we are going to determine the \emph{ $\pi^0$ decay photon
emergent spectrum}.\\
Let us consider the formation of a neutral, light, $\pi^0$ meson.
Light mesons, like $\pi^0$, represent very likely decay products
for heavy hadrons. Furthermore, the preferential
decay channel, $\pi^0\longrightarrow \gamma + \gamma$, produces two photons
which can be easily detected.\\
Hadronization is an intrinsically non-perturbative effect.
As such it is difficult  to describe in the framework of $QCD$.
Our ignorance about color non-perturbative dynamics can be parametrized in terms
of the so-called \emph{hadronization function}.
For a neutral pion the hadronization function reads \cite{cline:num}:

\begin{equation}\label{eqn:pion}
\frac{dg_{j\pi}(Q,E_\pi)}{dE_\pi}=\frac{15}{16}\sqrt{\frac{Q}{E_\pi^3}}
\left(1-\frac{E_\pi}{Q}\right)^2\; ,
\end{equation}
where $g_{j\pi}(Q,E_\pi)$ denotes the number of  $\pi^0$ produced in the
 energy range  $[E_\pi,E_\pi+dE_\pi]$ by an energy $Q$ parton of the
 $j$-th kind. Thus, the flux of neutral pions emitted by a black hole,
 per  unit time, in the energy range $[E_\pi,E_\pi+dE_\pi]$ can be written as

\begin{equation}\label{eqn:pionflux}
\frac{dN_\pi}{dE_\pi dt}= \sum_j \int_{E_\pi}^\infty dQ \frac{dg_{j\pi}(Q\ ,
E)}{dE_\pi}\,\frac{dN_j}{dQdt}\; ,
\end{equation}
where the sum runs over all the parton species relevant to neutral pion
formation.\\
As mentioned above, we can follow two approaches to clarify the physical meaning
of the  term $dN_j/dQ\, dt$:
\begin{itemize}
\item \emph{Direct Hadronization}: $\frac{dN_j}{dQdt}$ denotes the  number
of $j$-th species partons emitted
near the black hole horizon, per unit time, in the energy range  between $Q$ 
and
 $Q+dQ$, according with a black body spectrum

\begin{equation}\label{eqn:flux}
\frac{dN_j}{dQ dt}= \frac{\sigma_{gb}^{(s)}(Q)}{\pi^2}\frac{Q^2}{e^{Q/T_{
BH}}-(-1)^{2s}}\; ,
\end{equation}
where $s$ is the parton spin, and $\sigma_{gb}^{(s)}$ is
the grey-body cross section (Figure \ref{fig:GBF}). Once emitted near the 
horizon,quarks and gluons freely  propagate toward infinity. However, when the 
relative distance becomes higher than the threshold value
$\Lambda_{QCD}^{-1}\sim 1$ fm, then hadronization
starts.

\item \emph{Hadronization after chromosphere formation}: differently from the
preceeding case, once emitted near the horizon,
quarks and gluons propagate toward infinity forming a chromosphere,
as discussed previously. Thus, $\frac{dN_j}{dQ\, dt}$ denotes the  flux
of the $j$-th species partons
near the outer boundary  of the chromosphere; this flux can be evaluated 
following the method shown in reference \cite{cline:num}.
\end{itemize}
Then, the number of emergent photons, per unit time, in the energy range
$[E_\gamma,E_\gamma+dE_\gamma]$ is given by
\begin{equation}\label{eqn:fluxfot}
\frac{dN_\gamma }{dE_\gamma dt}= \int_{E_0}^\infty dE_\pi \frac{dg_{\gamma\pi}
(E_\pi,E_\gamma)}{dE_\gamma}\,\frac{dN_\pi}{dE_\pi dt}\; ,
\end{equation}
where, $E_0=E_\gamma+\frac{m_\pi^2}{4E_\gamma}$ is the minimum pion energy
which is necessary to  produce a photon  with energy $E_\gamma$, and 
 $m_\pi$ is the $\pi^0$ mass.
%%%%%%%%%%%%%%%%%%
%\footnote{Dallo studio del decay $\pi^0\rightarrow \gamma\gamma$ si p=
%u=F2 dimostrare che le energie massima e minima di un fotone emesso da un p=
%ione di energia $E_\pi= \gamma m_\pi$ sono $E_\gamma= E_\pi(1\pm\sqrt{1-\=
%gamma^{-2}})/2$. Rielaborando questa relazione =E8 possibile dimostrare che=
% un fotone di fissata energia $E_\gamma$ pu=F2 essere emesso da pioni con e=
%nergia tale che $E_\pi\geq E_\gamma+m_\pi^2/4E_\gamma$.};
%%%%%%%%%%%%%%%%%%
The function  $\frac{dg_{\gamma\pi}(E_\pi , E_\gamma)}{dE_\gamma}$ is the
number
of  energy $E_\gamma$ photons   produced by the decay of a pion 
with energy $E_\pi$ in the laboratory frame:

\begin{equation}\label{eqn:npifot}
\frac{dg_{\gamma\pi}(E_\pi,E_\gamma)}{dE_\gamma}= \frac{2}{\sqrt{E_\pi^2-m_\pi^
2}}\; .
\end{equation}
By taking into account equations (\ref{eqn:pion}), (\ref{eqn:pionflux}),
(\ref{eqn:fluxfot}) and (\ref{eqn:npifot}) one obtains

\begin{equation}\label{eqn:fluxfottot}
\frac{dN_\gamma}{dE_\gamma dt}= \sum_j \int_{E_0}^\infty dE_\pi \frac{15}{
8E_\pi^{3/2}\sqrt{E_\pi^2-m_\pi^2}} \,
\int_{E_\pi}^\infty dQ \sqrt Q
\left(1-\frac{E_\pi}{Q}\right)^2 \,\frac{dN_j}{dQ dt}\; .
\end{equation}
This integral can be computed numerically once the
 partonic flux $\frac{dN_j}{dQ dt}$ is given according with
 either  of the  approaches discussed above.\\
Equation (\ref{eqn:fluxfottot}) provides the total  flux of photons which
can be experimentally detected; these photons are not emitted by the black hole
itself but come from the pion decay.
Indeed we must to them add the near horizon ``direct'' emission spectrum. 
However,
also the direct emission photons can lead to the formation of a photosphere
through electromagnetic interaction with $e^\pm$ and produce a modified
spectrum
(Figure \ref{fig:11}).
In what follows, we shall list some results obtained for photon emergent 
spectra.
\subsection{Hadronization ``Post-Chromosphere''}\label{sez:Hpost}

\begin{figure}[htb]
\begin{center}
\includegraphics[width= 7.6cm]{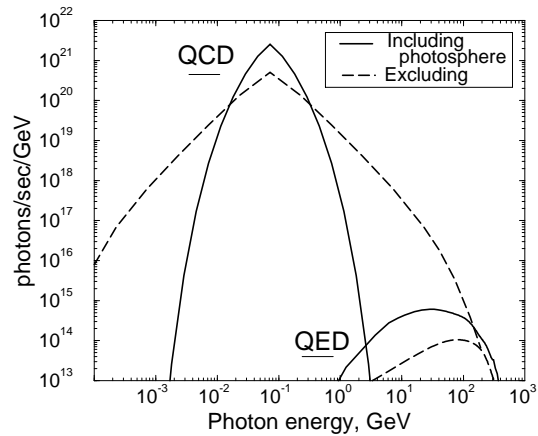}
\end{center}
\caption{\small Photon emission spectrum for a 4D Schwarzschild black hole
with temperature
$T_{BH}=50$ GeV. The   continuous line denote the spectrum obtained by taking
into
account the presence of photosphere (QED) and chromosphere (QCD).
The dashed lines denote either the  direct emission
spectrum, or the $\pi^0$ (QCD) decay spectrum following near horizon quark
hadronization (Reprinted figure with permission from \cite{cline:num}.
Copyright 1999 by The American Physical Society).}
\label{fig:19}
\end{figure}

\begin{figure}[htb]
\begin{center}
\includegraphics[width= 7.6cm]{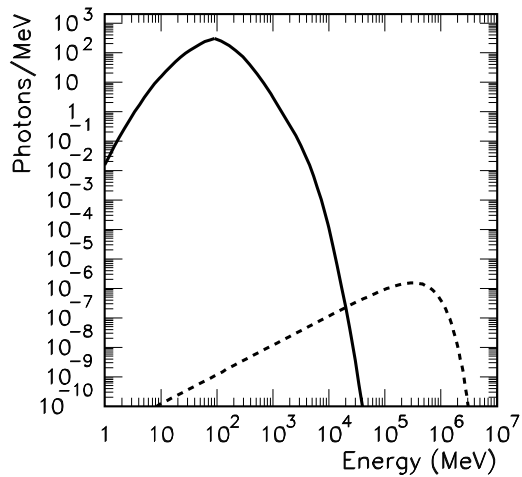}
\end{center}
\caption{\small Chromosphere emergent photon decay spectrum
 for a  Schwarzschild black hole in $D=10$
dimensions, with $M_*\simeq 1.3$ TeV and $M_{BH}/M_*= 5$. The
dashed line shows the direct emission spectrum (Reprinted figure with permission
from \cite{anchor:gold}. Copyright 2003 by The American Physical Society).}
\label{fig:anchor}
\end{figure}

In this subsection we shall focus on photon emergent spectra accounting
hadronization after chromosphere formation.\\
Figure \ref{fig:19} shows the total spectrum of photons emitted by
 a $4D$ Schwarzschild black hole with  temperature $T_{BH}=50$ GeV.
 Both direct emission photons (\underline{QED}) and $\pi^0$ (\underline{QCD})
 decay photons are displayed; each case is considered both when
 photo/chromosphere is present (solid curve), and when it is absent
(dashed curve). The whole photon spectrum is obtained  in either case
by summing the  solid and dashed lines.
From Figure \ref{fig:19} one sees that:
\begin{itemize}
\item  peak energy of $\pi^0$ decay photons corresponds to an
energy equal to $m_\pi/2$;  the   photon distribution is widened by
``Doppler effect'' as the pions do not decay at rest;
\item according with the previous remark, the photon spectrum  in the
absence of chromosphere is wider,  as the partons do not lose energy
in the chromosphere before hadronization;
\item the direct emission spectrum is picked  around
 $\sim 5T_{BH}$, while photosphere produces a larger number of photons with
smaller energy;
\item in the QED sector the photon peak  is many orders of magnitude smaller
than the photon peak in the QCD sector; as quarks and gluons dominate direct
emission, the QCD degrees of freedom, leading to pions, decaying into photons,
are many more than direct emission photons. Thus, we conclude that $\pi^0$ decay
photons dominate the emergent spectrum in Figure
\ref{fig:19}, both in presence and absence of  photo/chromosphere.
\end{itemize}
Figure \ref{fig:anchor} shows the decay photon spectrum in the presence of
chromosphere (solid line), and the direct emission spectrum (dashed line)
from a Schwarzschild black hole  in $D=10$ dimensions, for
$M_*\simeq 1.3$ TeV e $M_{BH}/M_*= 5$.
%%%%%%%%%%%%%%
%\footnote{Caso simile a quello considerato all'inizio della sezione \ref{se=
%z:schw} relativa allo studio della fase di Schwarzschild.};
%%%%%%%%%%%%%%
In this case $T_{BH}\simeq 200$ GeV  and the mean life is
$\tau_{BH}\simeq 4\times 10^{-27}$s.
A qualitative analysis of Figure \ref{fig:anchor} shows that the two curves
are peaked around  $m_\pi/2$ (solid line) and $\sim 5T_{BH}$ (dashed line),
that is, the involved physical processes
(direct emission, chromosphere development, hadronization and $\pi^0$ decay)
are spacetime dimension independent. This is consistent with the hypothesis
that elementary particle interactions are all localized on the spacetime 
$3$-brane and do not feel the presence of bulk extra dimensions.\\
We can estimate the number of $\pi^0$ decay photons emitted at the outer
boundary of the chromosphere to be roughly about $5\cdot 10^4$ photons
\cite{anchor:gold}. This number can be improved by adding the photons emerging
from the photosphere and the contributions from different hadronic decays.

%%%%%%%%%%%%%%%%%%%%%%%%%%%%%%%%%%
\subsection{Direct Hadronization}\label{sez:Hdirect}

\begin{figure}[htb]
\begin{center}
\includegraphics[angle=270,width=10cm]{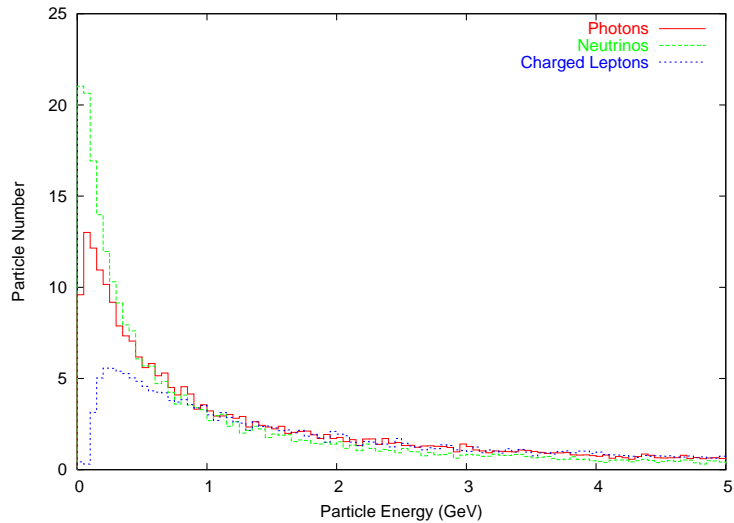}
\end{center}
\caption{\small Stable particle energy spectra for a $D=10$-dimensional
Schwarzschild black hole accounting direct hadronization.}
\label{fig:finalspt1}
\end{figure}

In this subsection we shall consider some results regarding final energy spectra
accounting direct hadronization.\\
In order to determine \emph{emergent spectra of stable species}
($e^\pm, (\nu\overline{\nu})_{e\mu\tau}$ and in particular $\gamma$),
production and decay of a $D$-dimensional Schwarzschild black hole
has been numerically performed by using black hole event generator
Charybdis v1.001 \cite{harris:art1}, while both the quark/gluon hadronization
and hadronic/leptonic decays have been numerically simulated by Pythia v6.227
\cite{pythia:62}.\\
In Table \ref{tab:parameters} we have reported some parameters obtained
for 100 generated events.
\begin{table}[htb]
\begin{center}
%\begin{size}
\begin{tabular}{|c||c|c|c|c|c|}
\hline{\rule[-3mm]{0mm}{8mm}}
$D$  & $\langle\,M_{BH}\,\rangle$ (TeV) & $\langle\,T_{BH}\,\rangle$ (GeV) & 
$\langle\,N_i\,\rangle$ & $\langle\,N_f\,\rangle$ & $\langle\,P\,\rangle$\\[2pt]
\hline{\rule[-3mm]{0mm}{8mm}}
6  &10.345 ($\pm$0.030)& 140.20 ($\pm$0.13) & 19.03 ($\pm$0.23)
& 1483 ($\pm$26) & 78.4 ($\pm$1.4)\\[4pt]
7  &10.279 ($\pm$0.030)& 235.62 ($\pm$0.17) & 13.06 ($\pm$0.22)
& 1244 ($\pm$24) & 96.4 ($\pm$1.9)\\[4pt]
8  &10.31  ($\pm$0.03) & 328.43 ($\pm$0.19) & 10.80 ($\pm$0.18)
& 1138 ($\pm$22) & 107.1 ($\pm$2.3)\\[4pt]
9  &10.299 ($\pm$0.027)& 416.28 ($\pm$0.18) &  9.57 ($\pm$0.18)
& 1028 ($\pm$22) & 109.5 ($\pm$2.5)\\[4pt]
10 &10.327 ($\pm$0.029)& 498.20 ($\pm$0.19) &  8.75 ($\pm$0.16)
& 1014 ($\pm$26) & 116.9 ($\pm$2.7)\\[4pt]
11 &10.259 ($\pm$0.024)& 575.37 ($\pm$0.16) &  8.19 ($\pm$0.16)
&  958 ($\pm$22) & 120 ($\pm$3)\\[4pt]
\hline
\end{tabular}
%\end{size}
\end{center}
\caption{\small Direct hadronization parameters obtained by 100 generated 
events. $\langle\,M_{BH}\,\rangle$ and $\langle\,T_{BH}\,\rangle$ are the 
average mass and temperature of a Schwarzschild black hole produced by Charybdis 
generator; $\langle\,N_i\,\rangle$ is the average number of particles directly 
emitted by black hole according to the Charybdis simulation, while 
$\langle\,N_f\,\rangle$ is the average number of emergent stable particles after 
Pythia simulated direct hadronization. $\langle\,P\,\rangle$ is the average 
production factor, i.e. the number of emergent stable particles following from 
each particle directly emitted by the black hole.}
\label{tab:parameters}
\end{table}
For different spacetime dimensions $D$,
the simulations show that the average black hole mass $\langle\,M_{BH}\,\rangle$
is approximatively costant, while the average temperature
$\langle\,T_{BH}\,\rangle$ grows; as a consequence less and less particles are
emitted out of the event horizon but they have more and more energy. Thus, the
average number of particles directly emitted by black hole
$\langle\,N_i\,\rangle$ and the average number of emergent stable particles
$\langle\,N_f\,\rangle$ decrease with $D$, while the average number of emergent
stable particles following from each particle directly emitted by the black hole
$\langle\,P\,\rangle$ is an increasing function of $D$. Accounting direct
hadronization, we can observe that black hole decay has very high multiplicity,
i.e. a black hole emits a great number of stable particles of O($10^3$).\\
In Figure \ref{fig:finalspt1} we show emergent energy spectra obtained from
numerical simulations.

\noindent
We notice that the final emergent spectra are dominated by photons and neutrinos.
Since future collider detectors are not tuned to capture neutrino signals,
a good signature could be missing energy or momentum;
thus, in Table \ref{tab:miss} we show the average missing energy
$\langle\,\emiss\,\rangle$ and the average missing transverse momentum
$\langle\,\ptmiss_T\,\rangle$ for several $D$.

\begin{table}[htb]
\begin{center}
%\begin{size}
\begin{tabular}{|c||c|c|c|}
\hline{\rule[-3mm]{0mm}{8mm}}
$D$ & 6 & 8 & 10 \\[2pt]
\hline{\rule[-3mm]{0mm}{8mm}}
$\langle\,\emiss\,\rangle$ (TeV) &
2.65 ($\pm$0.07) & 2.63 ($\pm$0.08) & 2.55 ($\pm$0.10) \\[4pt]
$\langle\,\ptmiss_T\,\rangle$ (TeV) &
1.82 ($\pm$0.05) & 1.81 ($\pm$0.07) & 1.76 ($\pm$0.08) \\[4pt]
\hline
\end{tabular}
%\end{size}
\end{center}
\caption{\small Average missing energy and average missing transverse momentum
for a $D$-dimensional Schwarzschild black hole accounting direct hadronization.}
\label{tab:miss}
\end{table}
\begin{figure}[htb]
\begin{center}
\includegraphics[angle=270,width=10cm]{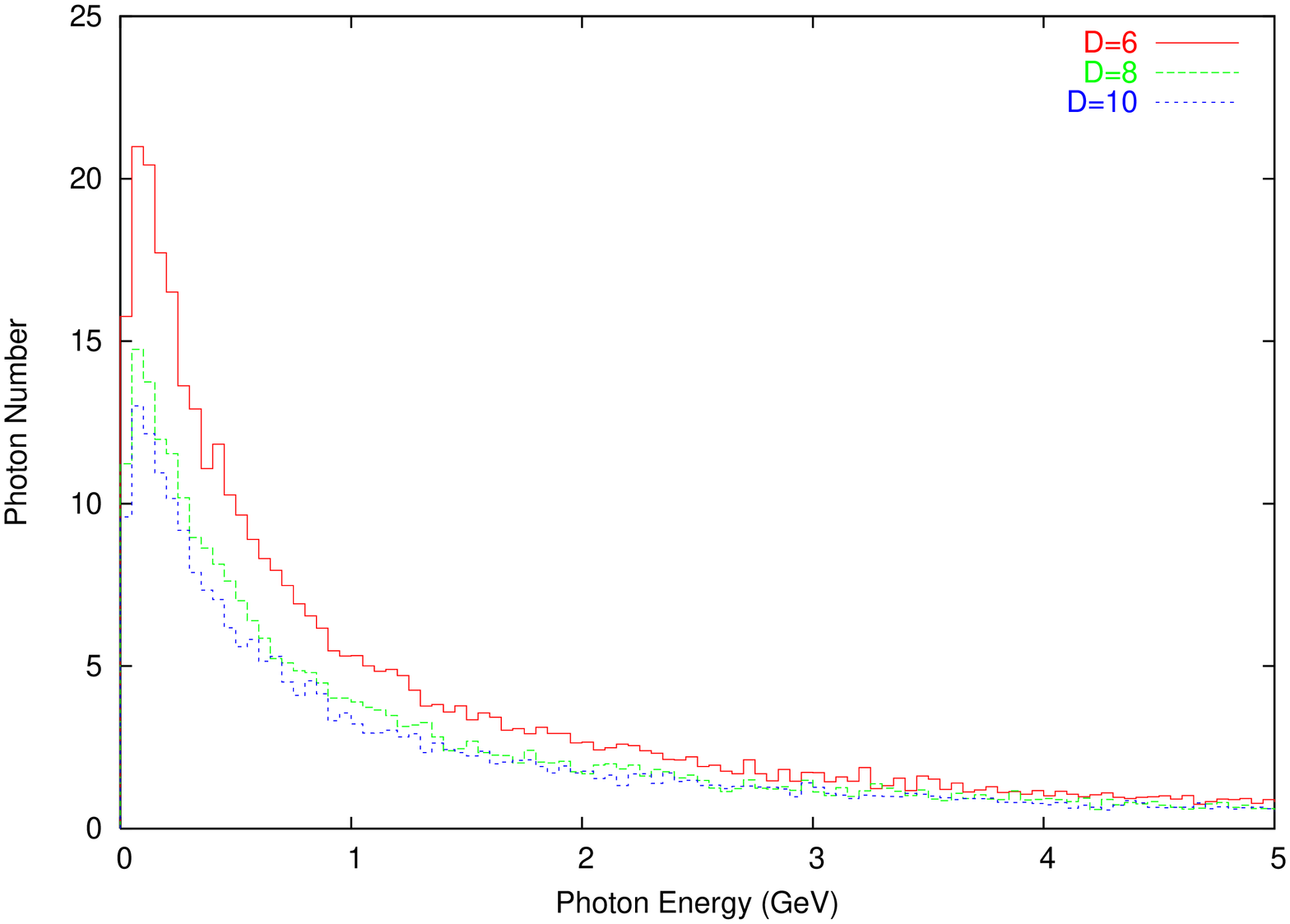}
\end{center}
\caption{\small Photon energy spectra for a $D$-dimensional Schwarzschild
black hole accounting direct hadronization.}
\label{fig:finalspt2}
\end{figure}
\noindent
Furthermore, in Figure \ref{fig:finalspt2} we show photon energy spectra; since
these spectra could be directly observed at future colliders, the whole picture
provides a possible experimental signature of TeV mini black hole evaporation.
%%%%%%%%%%%%%%%%%%%%%%%%%%%%%%%%%%%%%%%%%%%%%%%%%%%%%%%%%%%%%%%%%%%%%%%%%%%%%%%%
\section{Conclusions}\label{sez:conclu}

In this article we have reviewed TeV mini black hole decay in models with large
extra-dimensions, by taking into account not only the emission of particles
according to the Hawking mechanism (referred to as ``direct emission''),
but near horizon QCD/QED interactions, as well.\\
We have focused on higher dimensional Schwarzschild black hole decay
and we have observed that ``brane emission'' is parton dominated (see Table
\ref{tab:PF}). Since partons cannot be directly observed, we must take into
account fragmentation into hadrons; therefore, in order to
understand what kind of spectra we can expect to detect, we have reported
emergent photon spectra (Figures \ref{fig:19}, \ref{fig:anchor}, 
\ref{fig:finalspt1} and \ref{fig:finalspt2}), both in the case of more realistic 
near horizon QCD interactions (parton bremsstrahlung/pair production and after 
that fragmentation into hadrons) and in the case of ``direct 
hadronization''. Thus, one finds that final emergent spectra, to be measured by
an asymptotic observer, are dominated by hadrons and their decay products, 
mainly neutrinos and photons, both in presence and absence of photo/chromosphere.\\
In the latter case, we have reported some results obtained
by using Charybdis/Pythia event generator package: black hole decay event is
characterized by a large multiplicity, as high as $10^3$ (see Table
\ref{tab:parameters}) and a large missing energy and missing transverse
momentum, as high as TeV (see Table \ref{tab:miss}). The whole picture provides
a possible experimental signature at future colliders, as LHC and beyond.\\
However, much work has still to be done to obtain a phenomenologically
reliable signature at collider experiments. For instance, recoil effects of the
produced black hole \cite{tanaka:esc}, or the influence of Planck phase on the
experimental signature (\cite{hossen:due}, \cite{hossen:tre}) remain still to be
accounted for. In the latter case, according to several theoretical frameworks,
it has been argued that the final stage of black hole decay is characterized by
a ``remnant'' formation, i.e. either the black hole temperature abruptly drops
to zero \cite{adler:gup} or increases up to a maximum temperature and then
continuosly approaches an extremal, degenerate configuration at a finite black 
hole mass (\cite{nico:nocom},\cite{euro:nocom1},\cite{euro:nocom2}). The effects 
of this remnant formation on black hole evaporation have been investigated in 
\cite{hossen:tre}. The main result is that black hole emits a larger number of
Standard Model particles with a lower average energy and transverse momentum 
than in the case of total evaporation; more in detail, the total trasverse 
momentum ($\sum p_T$) is lowered by a quantity of the order of the remnant mass.\\
In conclusion, the main feature of black hole decay is the missing of
energy or transverse momentum. When a remnant is left after evaporation the
missing transverse momentum is of the order of the remnant mass. The big
acceptance of the detecting device at LHC will enable a complete event
reconstruction and to determine the missing energy. A precise estimate of the
missing energy in the framework of a specific regular black hole model is
currently under investigation by the authors and the results will be reported
elsewhere.

\vspace{1cm}

%%%%%%%%%%%%%%%%%%%%%%%%%%%%%%%%%%%%%%%%%%%%%%%%%%%%%%%%%%%%
%BIBLIOGRAFY (from file art.bib/bbl)
%%%%%%%%%%%%%%%%%%%%%%%%%%%%%%%%%%%%%%%%%%%%%%%%%%%%%%%%%%%%
%\noindent{\bf References}\\
{\footnotesize
\newcommand{\ç}{\"}
\bibliographystyle{unsrt}
\bibliography{art}
}
%%%%%%%%%%%%%%%%%%%%%%%%%%%%%%%%%%%%%%%%%%%%%%%%%%%%%%%%%%%%
\end{document}